# ANOTHER LOOK AT GAUSSIAN ISOCURVATURE HOT DARK MATTER MODELS FOR LARGE-SCALE STRUCTURE

Andrew A. de Laix [1]
and
Robert J. Scherrer [2]

ABSTRACT

We examine Gaussian isocurvature hot dark matter (massive neutrino) models for large-scale structure in which the initial density perturbations are produced in the baryons with a power–law spectrum $P_B(k) = Ak^{n_B}$. We calculate the linearly-evolved power spectrum and cosmic microwave fluctuations. We find that models with only isocurvature perturbations are inconsistent with observations of damped Ly$\alpha$ systems and COBE constraints on the power index. However, models which contain a mixture of adiabatic and isocurvature perturbations can be made consistent with COBE, galaxy surveys and damped Ly$\alpha$ systems. Isocurvature hot dark matter models also produce a bias between baryons and neutrinos even in the linear regime. We find that this "natural bias" can increase the baryon fraction in small scale objects like damped Ly$\alpha$ systems, but it has no effect on cluster scales.

---

[1]Department of Physics, Case Western Reserve University, Cleveland, OH 44106-7079
[2]Department of Physics, The Ohio State University, Columbus, OH 43210

# 1 Introduction

One candidate for the dominant matter in the universe is the massive neutrino, with a mass in the eV range. Since a neutrino with a mass in this range is still relativistic when structure formation begins at the epoch of equal matter and radiation, massive neutrinos act as hot dark matter (HDM) for the purposes of structure formation. Adiabatic (curvature) hot dark matter models with a Harrison-Zel'dovich power spectrum suffer from a well-known lack of power on small scales, due to the free-streaming of the neutrinos (White, Frenk, and Davis 1983; Kaiser 1983; White, Davis, and Frenk 1984; for an opposing point of view, see Melott 1985). This problem can be ameliorated by beginning with isocurvature fluctuations, either Gaussian fluctuations in the baryon component (the subject of this paper), or non-Gaussian isocurvature fluctuations due to some relic "seed" object. In these models, the isocurvature fluctuations preserve small scale perturbations until the neutrino streaming length becomes small enough to allow perturbation growth on such small scales, significantly enhancing the power on these scales.

The non-Gaussian HDM seed models have been investigated in great detail in recent years. Models with HDM and cosmic strings (Bertschinger & Watts 1988; Scherrer, Melott, & Bertschinger 1989; Albrecht & Stebbins 1992) or HDM and generic "seeds" (Villumsen, Scherrer & Bertschinger 1991; Gratsias et al. 1993) can produce much more power on small scales than adiabatic HDM and can give much better agreement with the observations. Hot dark matter with global texture (Cen et al. 1991) more closely resembles adiabatic HDM, because the texture unwinds on a relatively short time scale and does not remain around to seed structure at late times.

Here we examine Gaussian isocurvature HDM (IHDM) models, in which the fluctuations are generated initially in the baryon component. Gaussian models have the advantage of simplicity; the initial power spectrum totally specifies the model. Gaussian isocurvature perturbations are not as well-motivated as adiabatic perturbations, which arise naturally in the context of inflation. However, a number of models have been proposed which can give rise to isocurvature fluctuations in the baryons. For example, isocurvature fluctuations in the baryons can be produced by inhomogeneous baryogenesis (Turner, Cohen, & Kaplan 1989; Yokoyama & Suto 1991 and references therein), or by primordial magnetic fields (Kim, Olinto, & Rosner 1995). IHDM models were apparently first proposed by Peebles (1983) and inves-tigated more recently by Sugiyama, Sasaki, & Tomita (1989). In light of the stringent constraints which the COBE results place on any model for large-scale structure, this is an opportune time to re-examine these models.

In the next section, we discuss the initial conditions for the IHDM model, and calculate the cosmic microwave fluctuations and power spectra for isocurvature HDM models with an initial power-law fluctuation spectrum, normalizing to the COBE results. In section 3, we examine mixed isocurvature and adiabatic HDM models (MHDM), in which the former component dominates on small scales and the latter on large scales. In section 4 we discuss the data from the cosmic microwave background (CBR) and structure formation with which we will constrain our models. Finally, in section 5 we investigate a unique characteristic of isocurvature HDM models: such models naturally produce a bias between baryons and HDM even in the linear regime. In section 6 we present the results of our fits to the data. Our conclusions are summarized in section 7. We find that models with only isocurvature perturbations and an initial power law spectrum are ruled out by the existence of damped Ly$\alpha$ systems, but models with a mixture of adiabatic and isocurvature perturbations can fit the data; in particular, such models with white noise baryon fluctuations give an acceptable fit.

## 2 Power spectra and microwave fluctuations

We assume an $\Omega = 1$ universe dominated by a single massive neutrino with $m_\nu = 93(1 - \Omega_B)h^2$, where $h$ is the Hubble parameter in units of 100 km sec$^{-1}$ Mpc$^{-1}$ and $\Omega_B$ is the baryon fraction. (This differs from the models examined by Sugiyama et al. (1989), who assumed three massive neutrinos of equal mass). We take $h = 0.5$, consistent with universal age constraints, and assume $\Omega_B = 0.0125h^{-2}$, consistent with primordial nucleosynthesis (Walker *et al.* 1991).

The initial isocurvature fluctuations in the baryons are assumed to have Gaussian statistics and initial power spectrum $P_B(k)$. Because the fluctuations are isocurvature, the photons and neutrinos initially have compensating fluctuations which cancel the baryon fluctuations to give a small net perturbation (for a detailed analysis see Schaefer and de Laix 1995). For simplicity, we take $P_B(k)$ to have a power-law initial spectrum:

$$P_B(k) = Ak^{n_B}, \qquad (1)$$

where the amplitude $A$ and power index $n_B$ are to be fit to the observations. We calculate the final processed linear power spectrum $P(k)$ using the formalism of Schaefer and de Laix. For each component, this involves solving the linearly perturbed Boltzmann equation given by

$$\begin{aligned}\dot{\mathcal{F}} &+ ik\mu\frac{v}{q}\mathcal{F} + \frac{df}{dv}\left[\frac{v}{1+w}\frac{\dot{a}}{a}\left(c_s^2\Delta + w\Gamma - \frac{2}{3}w\Pi\right)\right]\\ &+i\mu\frac{df}{dv}\left[\frac{3q}{2k}\left[\frac{\dot{a}}{a}\right]^2(\Delta + 2w\Pi) + \frac{\dot{a}}{a}\frac{v^2}{q}V\right] = 0.\end{aligned} \quad (2)$$

Here, $\mathcal{F}$ is the gauge invariant perturbation in the phase space density $f$, where $\mathcal{F}$ is a function of the comoving momentum $\vec{v}$ and wave vector $\vec{k}$. $\dot{\mathcal{F}}$ is the derivative with respect to conformal time, $\mu$ is the cosine of the angle between $\vec{k}$ and $\vec{v}$ and $q$ is the comoving energy. The variables $\Delta, V$ and $\Pi$ are moments of the phase space density representing the energy density, velocity and anisotropic pressure fluctuations respectively (see Schaefer and de Laix 1995 for more details).

Asymptotic limits are easy to calculate: on scales larger than the horizon at equal matter and radiation, the usual isocurvature suppression gives $P(k) \propto k^{n_B+4}$, while in the limit of large $k$, the free-streaming of the neutrinos produces a power spectrum which asymptotes to $P(k) \propto k^{n_B-4}$ (Villumsen et al. 1991). What are reasonable values for $n_B$? If the process which produced the fluctuations in the baryons was uncorrelated on scales larger than the horizon when the perturbations were generated, then we expect on those larger scales a white-noise spectrum, for which $n_B = 0$. This spectrum is the Gaussian limit of the seeded hot dark matter model examined by Villumsen et al. (1991), who found that the model produced acceptable results on small scales. The limit $n_B = -3$ corresponds to the isocurvature Zel'dovich power spectrum (one of the two cases examined by Sugiyama *et al.* 1989, who also investigated $n_B = -1$). We will leave $n_B$ a free parameter to be fixed with respect to the observations contained in section 4.

First let us consider the baryon transfer function $T_B(k)$ defined by $T_B(k) = \Delta_B/\Delta_{B0}$ where $\Delta_{B0}$ and $\Delta_B$ are the initial and evolved perturbations respectively. Thus, $P_B(k) \propto T_B(k)^2 k^{n_B}$. We are left with the freedom to choose a convenient normalization for $T_B(k)$. It is easily verified that the growth factor for baryon fluctuations outside the horizon is the same regardless of scale, making the baryons a convenient reference point for defining transfer functions. We thus can choose the normalization such that $T_B(k) = 1$ in the limit of small $k$, while the transfer functions for the total perturbation and the HDM are defined in proportion to $T_B(k)$. This is illustrated in figure 1 which shows a plot of the linear transfer functions for the baryon, HDM and total perturbations evolved linearly to the present. For $k$ given in units of Mpc, the total linear transfer function is well fit by (better that 1%):

$$\begin{aligned} T(k) &= \frac{a_9 k^2}{|1 + a_1 k + a_2 k^2 + a_3 k^3 + a_4 k^4 + a_5 k^5 + a_6 k^6 + a_7 k^7|^{a_8}} \\ &+ a_{10} \left(1 + \frac{a_{11}}{k} + \frac{a_{12}}{k^2}\right)^{a_{13}} \exp\left(\frac{-a_{14}}{k}\right), \end{aligned} \quad (3)$$

where the values for the coefficients $a_i$ are given in table 1 for redshifts $z = 0$, 3.2, 4 and 9. Similarly one may fit the baryon transfer function with

$$\begin{aligned} T_B(k) &= \frac{a_9 \left(1 + a_{15} k + a_{16} k^2\right)}{|1 + a_1 k + a_2 k^2 + a_3 k^3 + a_4 k^4 + a_5 k^5 + a_6 k^6 + a_7 k^7|^{a_8}} \\ &+ a_{10} \left(1 + \frac{a_{11}}{k} + \frac{a_{12}}{k^2}\right)^{a_{13}} \exp\left(\frac{-a_{14}}{k}\right), \end{aligned} \quad (4)$$

where the coefficients are given in table 2.

With these two transfer functions, one may derive the results for the hot component using the formula

$$T_H(k) = \frac{\Omega_T}{\Omega_H} T(k) - \frac{\Omega_B}{\Omega_H} T_B(k), \quad (5)$$

where the $\Omega$'s are ratios of the energy density of the component to the closure density. It is important for the reader to realize that these transfer functions do not necessarily scale in any simple way with redshift, unlike, for example, those produced by adiabatic CDM fluctuations which scale as $(1+z)^{-1}$ in the matter dominated epoch. To illustrate this point, we have plotted in figure 2 the transfer functions for the baryons, hot component and total respectively evaluated at redshifts of $1 + z = 1, 5, 10$ and $20$. For the case of the total density perturbation, it is often adequate to just scale the transfer function by $(1+z)^{-1}$ for moderate redshifts. However, one can see clearly from figure 2 that such is not the case for baryons.

We would now like to turn to the effects IHDM will have on the cosmic microwave background. Because isocurvature perturbations produce only small fluctuations on super–horizon scales, the causes of the fluctuations in the cosmic microwave background are distinctly different from those produced by adiabatic models. For the case of adiabatic perturbations on large angular scales, like those probed by COBE, microwave distortions are generated by the Sachs–Wolfe effect, *i.e.* by the variations in the gravitational potential induced by density fluctuations (see *e.g.* Peebles 1993). Conversely, isocurvature perturbations produce only small total fluctuations so that it is the intrinsic fluctuation of the photon density which dominates, not the Sachs–Wolfe contribution. While the sources are different, the observed multipole moments, at least on large scales, have a similar $\ell$ dependence, where $\ell$ is the order of the multipole. However, the amplitude for $\Delta T/T$ given similarly–normalized power spectra will be different. One can see the similarity in the low order multipole moments by looking at the temperature fluctuations in Fourier space. For perturbations generated by the Sachs–Wolfe effect, the Fourier component of $\Delta T/T$ with wavenumber $k$ is related to the total density fluctuation $\Delta$ by $\Delta T/T \propto \Delta/k^2$. As it turns out, for IHDM, the perturbation in the photon density is also related to the total density by $\Delta_\gamma \propto \Delta T/T \propto \Delta/k^2$. Since $\Delta T/T$ scales similarly with $k$ for each case, both adiabatic and isocurvature models will produce similar multipole distributions on large scales, *i.e.* small $\ell$. On scales smaller than the horizon at recombination, the story is different. Here the microphysical interactions between the photons and baryons prior to decoupling determine the shape of the multipole distribution. The location and amplitude of Doppler peaks depends significantly on the velocity at which the perturbation in the photon baryon fluid crosses the horizon (see *e.g.* Hu and Sugiyama 1995); adiabatic perturbations cross the horizon with a significant velocity perturbation while isocurvature perturbations do not. In figure 3, we show a plot of the first 1500 coefficients to the multipole expansion, $a_\ell$, for IHDM models with power indices running from $n_B = -3$ to 0, where all coefficients are plotted as the ratio $a_\ell^2 \ell(\ell+1)/90 a_9^2$. For comparison, we also show the results for adiabatic HDM. The $n_B = -3$ IHDM model is comparable to the HDM plot as they are both produced by initially scale invariant spectra, and one can see the similarities on large scales (low numbered moments) between adiabatic HDM and IHDM with $n_B = -3$. One can also see that the Doppler peaks for adiabatic and isocurvature models are out of phase, a result anticipated by Hu and Sugiyama (1995).

# 3 Mixed adiabatic and isocurvature models

For the sake of completeness, we would like also to consider models which add an adiabatic (i.e. curvature) component to the IHDM model. One can find motivation for this approach from inflation. There are many compelling arguments in favor of inflation (Guth 1981), and it naturally leads to adiabatic fluctuations with a $k^1$ power spectrum, regardless of the presence of isocurvature fluctuations. In such a model, the isocurvature fluctuations must be produced either during or after inflation to avoid being erased by inflation; many models for producing isocurvature fluctuations during inflation have been proposed (Barrow & Turner 1981; Bond, Kolb & Silk 1982; Yoshimura 1983; Linde 1985; Seckel & Turner 1985; Fukugita & Rubakov 1986; Kofman 1986; Kofman & Linde 1987; Kofman & Pogosyan 1988; Turner, Cohen & Kaplan 1989). Gratsias et al. (1993) suggested the addition of adiabatic Zel'dovich fluctuations to the seeded hot dark matter model in order to enhance the power on large scales. Lilje (1990) argued in the opposite direction, pointing out that adiabatic Zel'dovich HDM produces acceptable power on large scales but needs an additional fluctuation component on small scales.

We assume, then, an adiabatic component with an initial Zel'dovich power spectrum, giving the present power spectrum

$$P_A(k) = A_1 T_A^2(k) k. \qquad (6)$$

Our numerical analysis has shown that for an adiabatic power spectrum defined by $P_A(k) = 1.2 \times 10^7 A_1 k T_A^2(k)$ Mpc$^3$ that $a_{9\ A} = a_{9\ COBE}$ when $A_1 = 1$. For convenience we have redefined $T_A(k)$ so when $A_1$ is equal to unity, we have a COBE normalized adiabatic spectrum. We add this to our present power-law isocurvature spectrum

$$P_I(k) = A_2 T^2(k) k^{n_B}. \qquad (7)$$

Recall that $T(k)$ is the total present isocurvature transfer function given by equation (3). We will further assume that the processes that produce the adiabatic and isocurvature fluctuations are statistically uncorrelated which eliminates any cross terms in the power spectrum. That is, the total power spectrum may be written as the sum of the adiabatic and isocurvature components, *i.e.* $P(k) = P_A(k) + P_I(k)$. We are left with a model completely described by three free parameters, $A_1$, $A_2$ and $n_B$. The constraints on these parameters are obtained from observational data which we shall discuss in the following section.

# 4 Limits From COBE and Structure Formation

For both IHDM and MHDM models we have a number of free parameters which we would like to constrain with observations. Fortunately, data is available on many different length scales to provide constraints. On the largest scales, near the horizon size, we have the microwave background fluctuations as measured by COBE. On cluster scales ($8h^{-1}$ Mpc) we have data from the QDOT *IRAS* galaxy survey, and on small scales ($0.13h^{-1}$ Mpc) we have observations of damped Ly$\alpha$ systems. To constrain our models with this diverse set of measurements we have elected to perform a $\chi^2$ minimization on all data simultaneously to determine which parameters best fit the data as a whole. From this we can establish a likelihood that a particular model can explain observations on many different length scales. So let us now consider each of the data sets in more detail.

We shall first turn to the COBE data. An analysis performed by Gorski *et al.* (1994) on the two year COBE data has shown for Sachs–Wolfe temperature fluctuations generated by a power law density spectrum, *i.e.* $P(k) \propto k^n$, that the ninth coefficient of the multipole expansion is a constant for a broad range of power indices. We will take the Gorski *et al.* value of $a_{9\ COBE} = 8.3 \pm 0.7\ \mu K/T_{\gamma 0}$ with $T_{\gamma 0} = 2.726\ \mu K$ as the the COBE normalization data point for our fits. Using our numerical simulations, we have determined an analytic fit for $a_9$ as a function of $n_B$ for the IHDM models. Note that for this calculation we have defined the power spectrum by $P(k) = A_2 k^{n_B} T^2(k)$ Mpc$^3$ where $k$ is given in units of Mpc$^{-1}$. The fit for $a_9$, valid for the range $n_B = -3$ to 2 is given by

$$\log_{10}\left(\frac{a_9^2}{A_2}\right) = \sum_{i=0}^{6} c_i n_B^i, \qquad (8)$$

where the coefficients $c_i$ are given in table 3.

Gorski *et al.* have also constrained the value of the power index using the COBE data. For IHDM models, that is models with no adiabatic contribution, this power index constraint translates into $n_B = -3.02 \pm 0.36$, so we shall include this in our fits of IHDM models. The COBE observations prefer a Harrison-Zel'dovich spectrum which is reflected in the constraint on $n_B$. For MHDM models, the large scale temperature fluctuations are produced by the adiabatic component which satisfies the power index constraint, so we will ignore it in this case.

To make any further constraints, we must turn to structure observations. For scales larger than $8h^{-1}$ Mpc, non-linear evolution is not significant, and any data that is available should correspond closely to the linear power spectrum. Such data is available from an analysis performed on the QDOT $IRAS$ survey (Efstathiou *et al.* 1990) by Feldman *et al.* (1994). They were able to estimate the power spectrum in the range of $k = 0.02$ to $0.2\ h$ Mpc, well within the linear regime. However, since the data on which they based their observations are from redshift surveys, they are not equivalent to real space measurements, and thus corrections must be applied to relate the observed results to theoretical power spectra. Redshift surveys tend to show greater clustering than is actually present (Kaiser 1987), so a correction factor is introduced such that

$$P(k) \to b_I^2 \left[1 + \frac{2}{3b_I} + \frac{1}{5b_I^2}\right] P(k). \qquad (9)$$

The parameter $b_I$ is a bias factor which represents the uncertainty to which the $IRAS$ galaxies trace the underlying mass distribution. Measurement of the large scale peculiar velocity field from the QDOT data performed by Kaiser *et al.* (1991) gives the value of $b_I = 1.16 \pm 0.21$. We shall allow $b_I$ to be a free parameter to be best fit by our models. A second correction is required on small scales since the peculiar velocities of the galaxies tend to overshadow the clustering effect (Peacock 1992). This can be accounted for by multiplying by a damping factor

$$P(k) \to \frac{\sqrt{\pi}}{2} \frac{\operatorname{erf}(kR_v)}{kR_v} P(k), \qquad (10)$$

where $R_v = 4.4h^{-1}$ Mpc.

Where the QDOT analysis yields information on large scales, the observation of damped Ly$\alpha$ systems yield constraints on small scales. Recent observations have detected such systems out to redshifts of $z \sim 4$ (Storrie–Lombardi *et al.* 1995). They estimate the density fraction of the neutral hydrogen in damped Ly$\alpha$ systems, $\Omega_{gas}$, for a flat ($\Omega = 1$) universe with $h = .5$ and find that $\Omega_{gas} = 2.2 \pm 0.4 \times 10^{-3}$ for $z = 4$ and $\Omega_{gas} = 3.3 \pm 0.6 \times 10^{-3}$ for $z = 3.2$. Data for smaller redshifts show strong indications of evolution, which tends to reduce $\Omega_{gas}$, so we will not consider this data. We will assume also that all the gas is neutral, ignoring the possibility that some might be ionized. Damped Ly$\alpha$ systems are optically dense, allowing only surface ionization, so this is not an unreasonable assumption.

The length scales corresponding to damped Ly$\alpha$ have undergone highly non–linear evolution, but we can estimate the Ly$\alpha$ mass function from the linear power spectrum using the Press–Schechter technique (Press & Schechter 1974; see *e.g.* Peebles 1993). In this formalism, the differential density fraction is given by

$$d\Omega = -\frac{\Omega_b}{1+Y}\sqrt{\frac{2}{\pi}} d\ln\sigma \ \nu_c e^{-\nu_c^2/2}, \qquad (11)$$

where $Y$ is the helium mass fraction, $\nu_c = \Delta_c/\sigma$ and $\Delta_c$ represents the critical collapse density. The density contrast $\sigma$ is the smoothed *rms* mass fluctuation given by

$$\sigma^2 = \frac{1}{2\pi^2} \int dk k^2 P(k,z) e^{-k^2 R^2}, \qquad (12)$$

where $R$ is the smoothing length and $P(k,z)$ is the power spectrum at redshift $z$. For damped Ly$\alpha$ systems, we take the minimum value for $R$ necessary to produce an observed cloud to be $R_{DLy\alpha} = 0.13h^{-1}$ Mpc (de Laix, Scherrer and Schaefer). This is our lower bound for integrating equation (11); we take an infinite upper bound. The value for $\Delta_c$ is uncertain because it depends on the details of the collapse. If one uses a simple spherical collapse model, one would derive a value of $\Delta_c = 1.69$; however, fits to hydrodynamical simulations have produced results as low as $\Delta_c = 1.33$ (see *e.g.* Klypin *et al* 1994). We shall consider both possibilities when calculating our model fits to damped Ly$\alpha$ systems.

## 5  Natural Bias

One of the unanswered questions in large-scale structure is the extent to which the baryon distribution traces the dark matter. This uncertainty is quantified in terms of the bias parameter $b$, defined by

$$b(R) = \sigma_B(R)/\sigma(R), \qquad (13)$$

where $\sigma(R)$ and $\sigma_B(R)$ are the smoothed *rms* density fluctuations for the total and baryon perturbations respectively. The possibility that $b \neq 1$ is usually attributed to poorly-understood processes which operate in the strongly nonlinear regime, particularly effects due to pressure which affect only the baryons. The IHDM model is unusual in that the distribution of baryons and HDM is naturally biased *in the linear regime*. This is due to the fact that when the neutrino free-streaming length is large, the neutrino perturbations on small scales are erased, while the baryon perturbations remain; we shall refer to this effect as "natural bias". However, as the neutrino free-streaming length drops, the combined neutrino and baryon perturbations can begin growing again, and this unbiased growing component eventually swamps the initial pure baryon fluctuation. Hence, we expect $b$ to be largest on small scales, and to be a decreasing function of time. This effect is obvious in the power spectrum derived by Sugiyama et al. (1989), who did not comment on it.

We can see from figure 1 that natural biasing can be significant on the scales of damped Ly$\alpha$ systems, $\sim 0.1\ h^{-1}$ Mpc. It is important then to see how this natural biasing will affect limits derived from damped Ly$\alpha$ clouds. One would like to know the ratio of $\rho_B/\rho$ in a collapsed cloud which, because of natural biasing, should not in principle be $\Omega_B$. We consider here only the natural bias in the linear density field, ignoring the non–linear effects. (Of course, this problem can only be completely resolved with hydrodynamic simulations, which are beyond the scope of this work.) With this restriction, we shall define the enhanced baryon fraction which we shall use in our Press-Schechter integral as

$$\frac{\rho_B}{\rho} = \Omega_B \left( \frac{1 + b\Delta_c}{1 + \Delta_c} \right). \tag{14}$$

This formula gives the ratio of baryon density to total in the linear regime for an object smoothed on a scale $R$ with an overdensity $\Delta_c$. One may be a bit suspicious that we have used the natural bias $b$ derived from the *rms* fluctuation when in fact we are only considering objects in highly over-dense regions, where the value of $b$ from natural bias might be quite different. To test this we have numerically generated Gaussian density fields for the baryonic and total perturbations smoothed over various scales for which natural biasing is significant. Our results indicate that, when considering only overdense regions with $\Delta > \Delta_c$ under the condition $\Delta_c/\sigma \gtrsim 1$, the ratio of $\Delta_B/\Delta$ is on average equal to $b$. The effect of natural bias then is to modify the Press-Schechter formula to give the following:

$$d\Omega_{gas} = -\frac{\Omega_b}{1+Y}\left(\frac{1+b\Delta_c}{1+\Delta_c}\right)\sqrt{\frac{2}{\pi}}d\ln\sigma\ \nu_c e^{-\nu_c^2/2}. \qquad (15)$$

We use this to estimate our natural biased density in damped L $\alpha$ systems. The value of $b$ is highly sensitive to $n_B$. For example, with $R = 0.1\ h^{-1}$ Mpc, we find that $b = 4.5$ for $n_B = 0$, while $b = 1.1$ for $n_B = -3$.

## 6 Results

We have proposed two classes of models and have presented the data with which we can constrain them, and we now present the results from our various fits. First let us consider the simpler of the two models, pure IHDM with a power–law power spectrum. Since we have no adiabatic contribution, we are left with three free parameters ($A_2$, $n_B$ and $b_I$). We find the best fit values by performing a $\chi^2$ fit to the data discussed in section 4, specifically COBE, the QDOT power spectrum and damped Ly$\alpha$ systems. The best fit $n_B$ is $\approx -2$, but this value of $n_B$ is a poor fit to the Ly$\alpha$ data. It produces so little power on small scales (asymptotically $P(k) \sim k^{-6}$) that no IHDM model consistent with COBE and QDOT can produce a significant density of damped Ly$\alpha$ systems at high redshift. Because of this lack of small-scale power, changing the value of $\Delta_c$ or including natural bias effects does not alter the fit.

Now let us add an adiabatic component to the isocurvature. Figure 4 presents the best fit values for $A_2$ and $n$ with all four parameters ($A_1$, $A_2$, $n_B$ and $b_I$) left free assuming no natural bias and a value for $\Delta_c = 1.69$. Also shown are the 68% and 95% confidence limits plotted in the $A_2 - n_B$ plane with $A_1$ and $b_I$ held fixed at their best fit values. The value of $A_1$ is fixed by COBE since the large scale microwave fluctuations are dominated by the adiabatic component for reasonable models. The adiabatic component also dominates the the power spectrum on large scales, so that it fixes the value of $b_I$ as well. The amplitude and power index ($A_2$ and $n_B$) of the isocurvature component are then determined by fitting the limits from damped Ly$\alpha$ systems but are restricted by the small scale results from QDOT as well. Figure 5 shows similar results as the previous figure accept that the value for $\Delta_c$ is now 1.33. Figure 6 shows the same plot as figure 4 with $\Delta_c = 1.69$ but now the natural bias effect discussed in section 5 is included. Finally, Figure 7 shows the results including the natural bias effect with $\Delta_c = 1.33$.

If we accept the lower bound on $\Delta_c = 1.33$ as reasonable, we see that white noise ($n_B = 0$) models are allowed at the 68% confidence limit for both the natural biased and unbiased cases. These models are of particular interest as they would arise from any process which is uncorrelated on scales larger than the horizon, and may be considered in some sense "natural".

Since we have considered all of the data simultaneously in our $\chi^2$ fits, it is important to examine each observation independently. First consider the COBE constraint. The normalization for the adiabatic HDM component, which determines the value for $a_9$, is well within the 2-$\sigma$ limit determined from COBE for both the natural biased and unbiased fits. Now consider QDOT. The large number of data points from the QDOT data set assures that any acceptable model should fit it well in a $\chi^2$ analysis, a fact that can be verified visually. Figure 8 shows a plot of the QDOT data along with the best fit IHDM model and the best fit MHDM model with $n_B = 0$ and natural bias. Also included for comparison is the cold+hot dark matter model (C+HDM) with 25% HDM and 75% CDM. It is normalized to COBE using the some $b_I$ as MHDM. The bias needed to fit the QDOT data is given by $b_I \approx 0.66$ for either the natural biased or unbiased fit. This is 2.4$\sigma$ from the mean estimate of $b_I = 1.16 \pm 0.21$ (Kaiser *et al.* 1991). Finally, consider the damped Ly$\alpha$ systems. We would like to see how the theoretical predictions for $\Omega_{gas}$ stand up to the observed data. For the best fit, unbiased MHDM model, estimates of the gas fraction are $\Omega_{gas} = 1.0 \times 10^{-3}, 2.4 \times 10^{-3}$ for $z = 4, 3.2$ respectively. That is a 3$\sigma$ difference for $z = 4$, the larger difference. With natural bias, the best fit gas fraction increases to $\Omega_{gas} = 1.3 \times 10^{-3}, 2.8 \times 10^{-3}$ for $z = 4, 3.2$ respectively. This gives a 2.5$\sigma$ difference at $z = 4$, a significant improvement over the unbiased model. For comparison, the C+HDM model with $\Delta_c = 1.33$ gives $\Omega_{gas} = 3.2 \times 10^{-3}, 5.6 \times 10^{-3}$ for $z = 4, 3.2$ respectively. C+HDM greatly overproduces damped Ly$\alpha$ systems, but this is not as bad as it seems. We chose an infinite upper scale for our Press–Schechter integral, but a finite upper bound, determined from the rotation of spiral galaxies which damped Ly$\alpha$ systems are believed to be the progenetors, may be more realistic (Mo & Miralda–Escudé 1994; Kaufman & Charlot 1994). One can find ways to reduce the observed number of systems, but it is more difficult to increase that number. The C+HDM model is probably a better fit. Finally, in figure 9 we present the multipole distribution for the best fit, natural bias model with a white noise ($n_B = 0$) input spectrum. The dashed line shows the isocurvature contribution, the dot–dashed the adiabatic contribution and the solid line shows the sum. The results show that on COBE scales, this MHDM model in indistinguishable from standard adiabatic models; however, for moments $\ell \geq 100$, there are significant deviations from pure adiabatic models which could be detected in future observations.

## 7 Conclusions

The results presented in section 6 give a clear picture of the types of models consistent with current observation. The most significant conclusion is that pure IHDM models are ruled out by the existence of high redshift damped Ly$\alpha$ systems. If we ignore these objects completely, IHDM is still lies on tenuous ground with regards to COBE, as the best fit power index of $n_B = -2.23$ lies outside the 95% confidence limit of the COBE power index constraint. One can reasonably reject IHDM with a power law initial spectrum as a viable model.

Conversely, MHDM should be considered a reasonable candidate. The largest strike against it is the unnaturally small value for $b_I$ required to fit QDOT. Measurements from galaxy redshift surveys like QDOT pose significant problems for $\Omega = 1$ models which do not contain a cosmological constant. It is difficult to reconcile COBE constraints on the power spectrum amplitude with $b_I = 1$ and QDOT. However, there is evidence that redshift surveys underestimate the large scale power. A recent analysis of the APM optical galaxy survey (Gaztañaga 1995) has yielded a value for $\sigma_{8h^{-1}Mpc} = 0.95 \pm 0.07$, significantly higher than those computed with $b_I = 1$ yielding $\sigma_{8h^{-1}Mpc} = 0.75 \pm 0.098$ (Peacock and Dodds 1994). For the best fit white noise, natural biased MHDM model, $\sigma_{8h^{-1}Mpc} = 1.1$, while the C+HDM model gives $\sigma_{8h^{-1}Mpc} = 0.91$.

MHDM models are also at the fringe when it comes to the observation of damped Ly$\alpha$ systems, especially at $z = 4$. However, there is again more breathing room than at first appears. Because of the statistical difficulties, we did not include the fact that the redshifts at which $\Omega_{gas}$ are observed are uncertain as well. The $1 - \sigma$ uncertainties quoted from Storrie-Lombardi *et al.* (1995) for the redshifts are $z = 4^{+0.8}_{-0.6}$ and $z = 3.2^{+0.4}_{-0.2}$. If the actual redshifts are lower than the mean value, the damped Ly$\alpha$ constraint would be significantly loosened.

While adiabatic + isocurvature models do not provide the best fit to the data currently available, they are not in obvious contradiction to any of the data. The C+HDM model offers a superior fit to both $QDOT$ and damped Ly$\alpha$ system data, but its superiority is not compelling. Thus, MHDM models produce a natural alternative to C+HDM models for large scale structure. The MHDM models have the advantage of requiring only a single dark matter particle (which is known to exist), but they require an additional mechanism to generate isocurvature perturbations.

# Figure Captions

1. The isocurvature hot dark matter transfer functions for the total (solid), baryon (dashed) and hot dark matter (dot–dashed) perturbations evolved linearly to the present. The normalization is chosen such that $T_B(k) = 1$ for small $k$.

2. The evolution of the linear isocurvature hot dark matter transfer functions in redshift: hot dark matter (top), baryons (middle) and total (bottom). Plots are shown for $1 + z = 1, 5, 10$ and $20$.

3. $\Delta T/T$ multipole ratio plots for isocurvature hot dark matter models $n_B = 0, -1, -2$ and $-3$ (solid curves) and for adiabatic hot dark matter (dashed curve).

4. The $A_2 - n$ best fit along with the 68% and 95% confidence limits for the mixed (adiabatic + isocurvature) hot dark matter model including no natural bias effect; $\Delta_c = 1.69$, $A_1 = 0.80$ and $b_I = 0.68$.

5. Same as 4 except $\Delta_c = 1.33$, $A_1 = 0.81$ and $b_I = 0.67$.

6. Same as 4 except now the effects of natural bias are included; $\Delta_c = 1.69$, $A_1 = 0.82$ and $b_I = 0.65$.

7. Same as 4 except now the effects of natural bias are included; $\Delta_c = 1.33$, $A_1 = 0.82$ and $b_I = 0.65$.

8. The QDOT power spectrum (Feldman *et al.* 1994) along with cold + hot dark matter (solid), isocurvature hot dark matter (dashed) and adiabatic + isocurvature hot dark matter (dot-dashed) power spectra.

9. The $\Delta T/T$ multipole moments for the best fit adiabatic + isocurvature hot dark matter model (including natural bias) with a white noise ($n_B = 0$) initial baryon power spectrum. The dashed curve shows the isocurvature contribution, the dot–dashed curve the adiabatic contribution and the solid curve the sum.

## Table Captions

1. Fitting coefficients for the isocurvature hot dark matter total transfer function $T(k)$ given at selected redshifts, where $k$ is given in units of Mpc$^{-1}$ and $h = 0.5$. Entries are written in the form $a(b) = a \times 10^b$.

2. Same as table 1 for the baryon transfer function $T_B(k)$.

3. Fitting coefficients for $a_9$ as a function of $n$. Entries are written in the form $a(b) = a \times 10^b$.

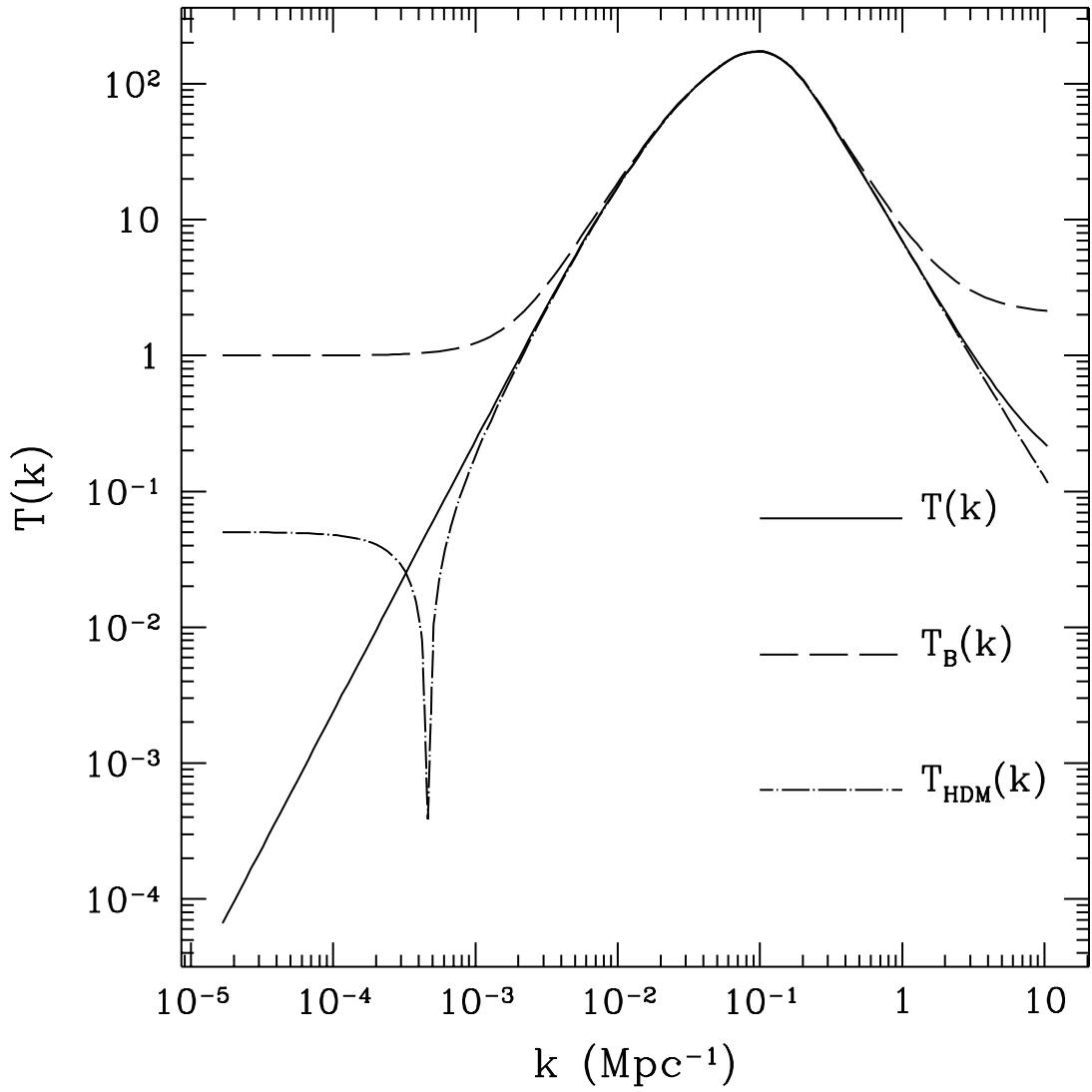

Figure 1:

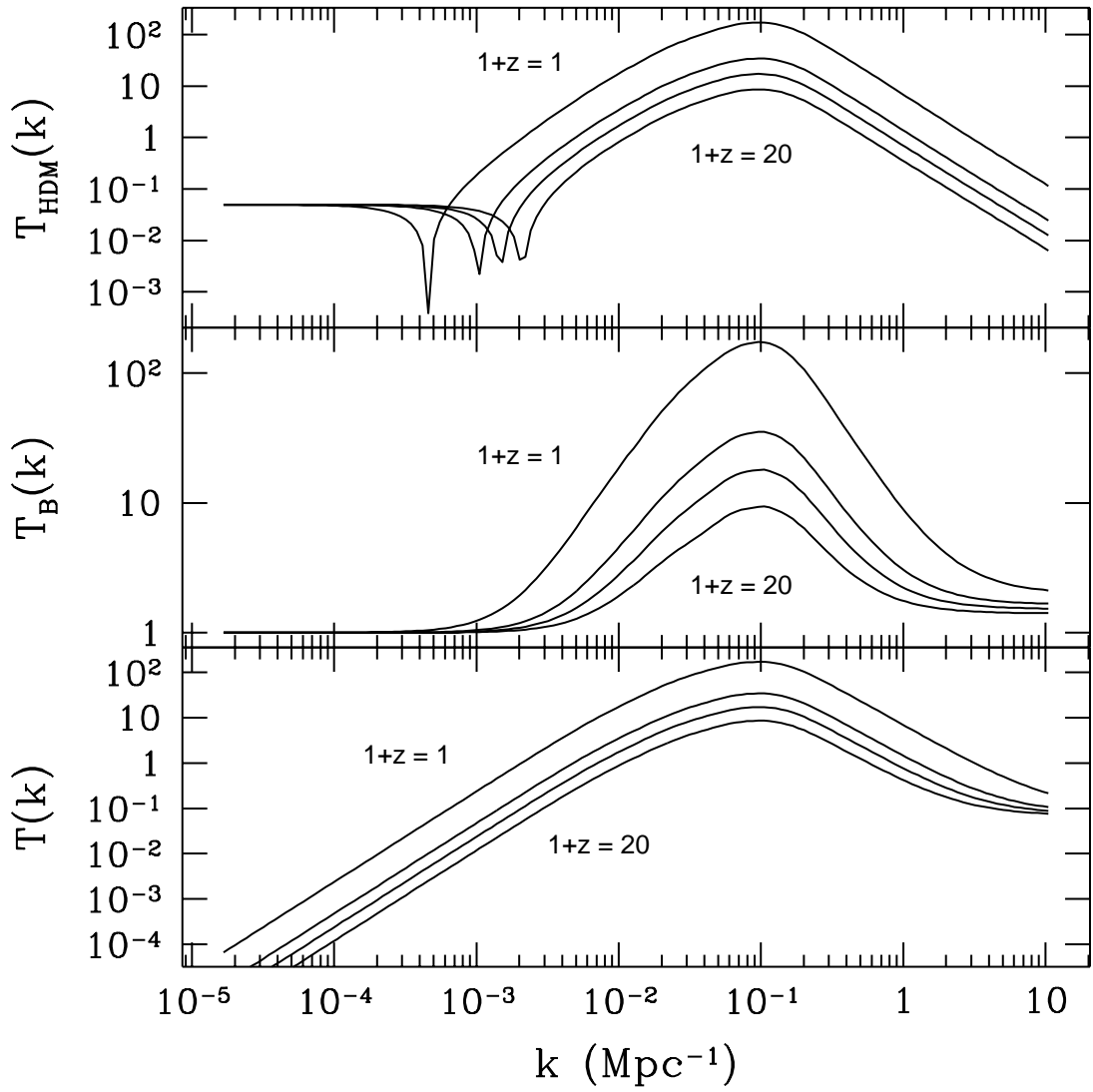

Figure 2:

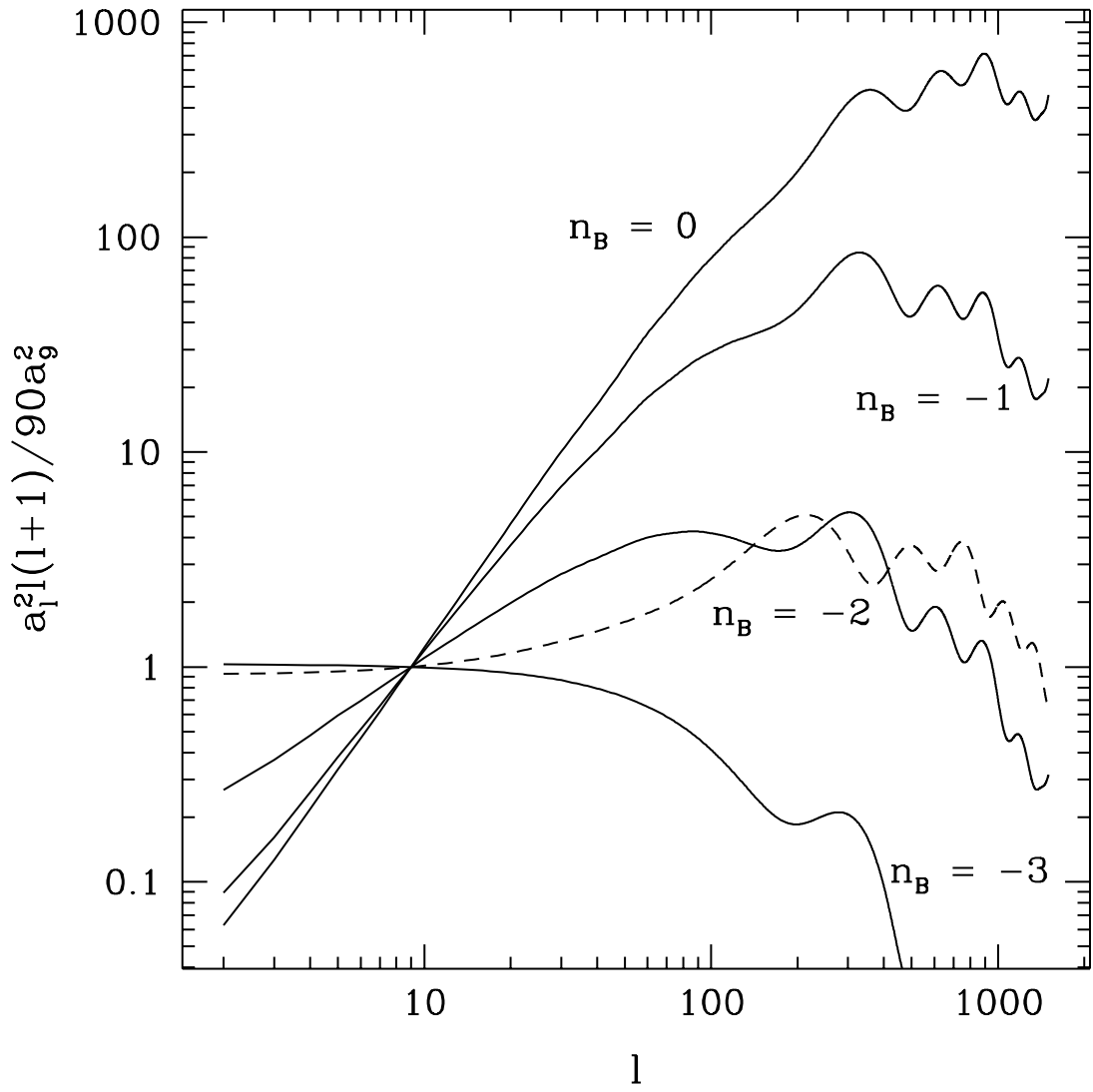

Figure 3:

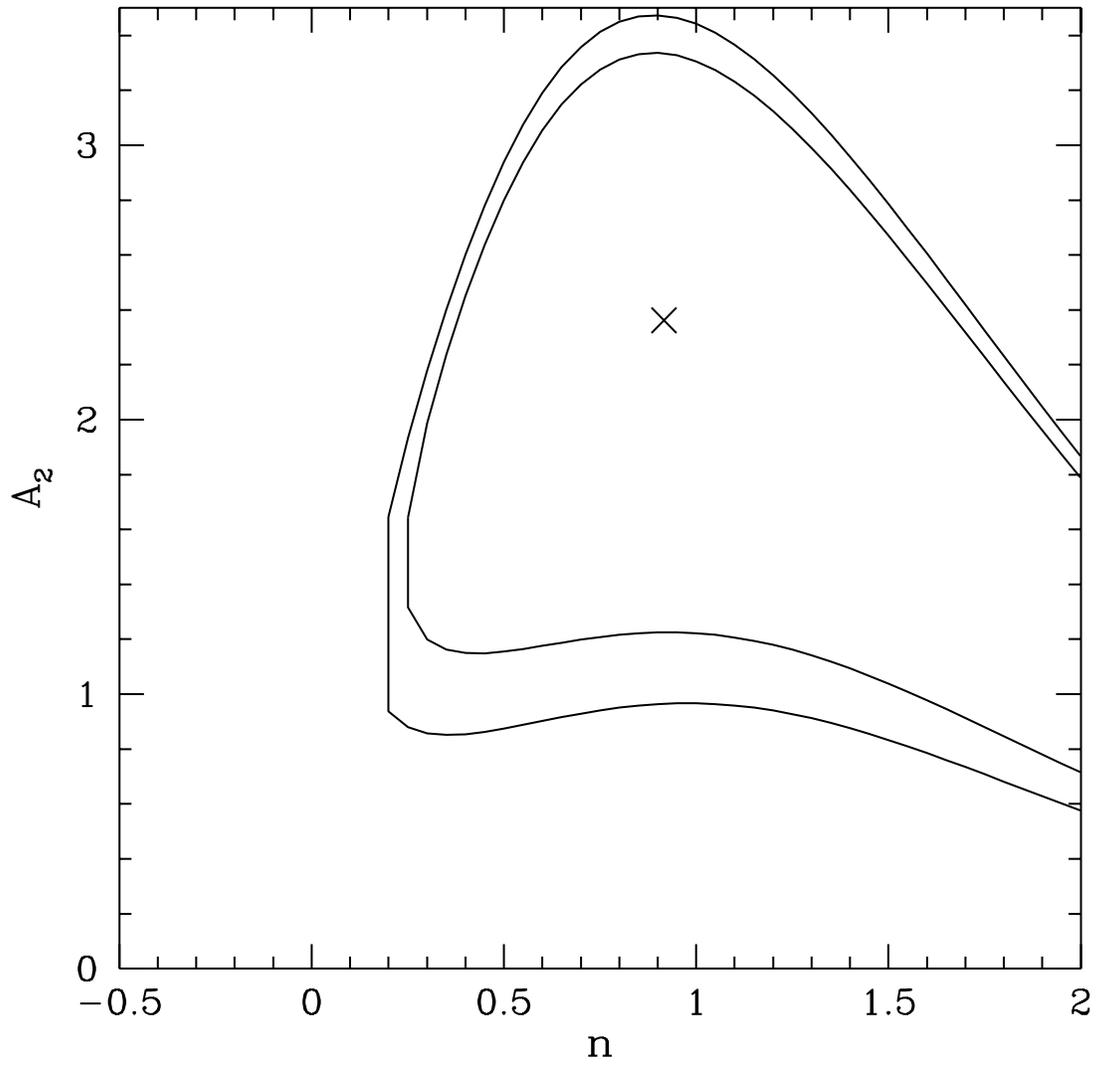

Figure 4:

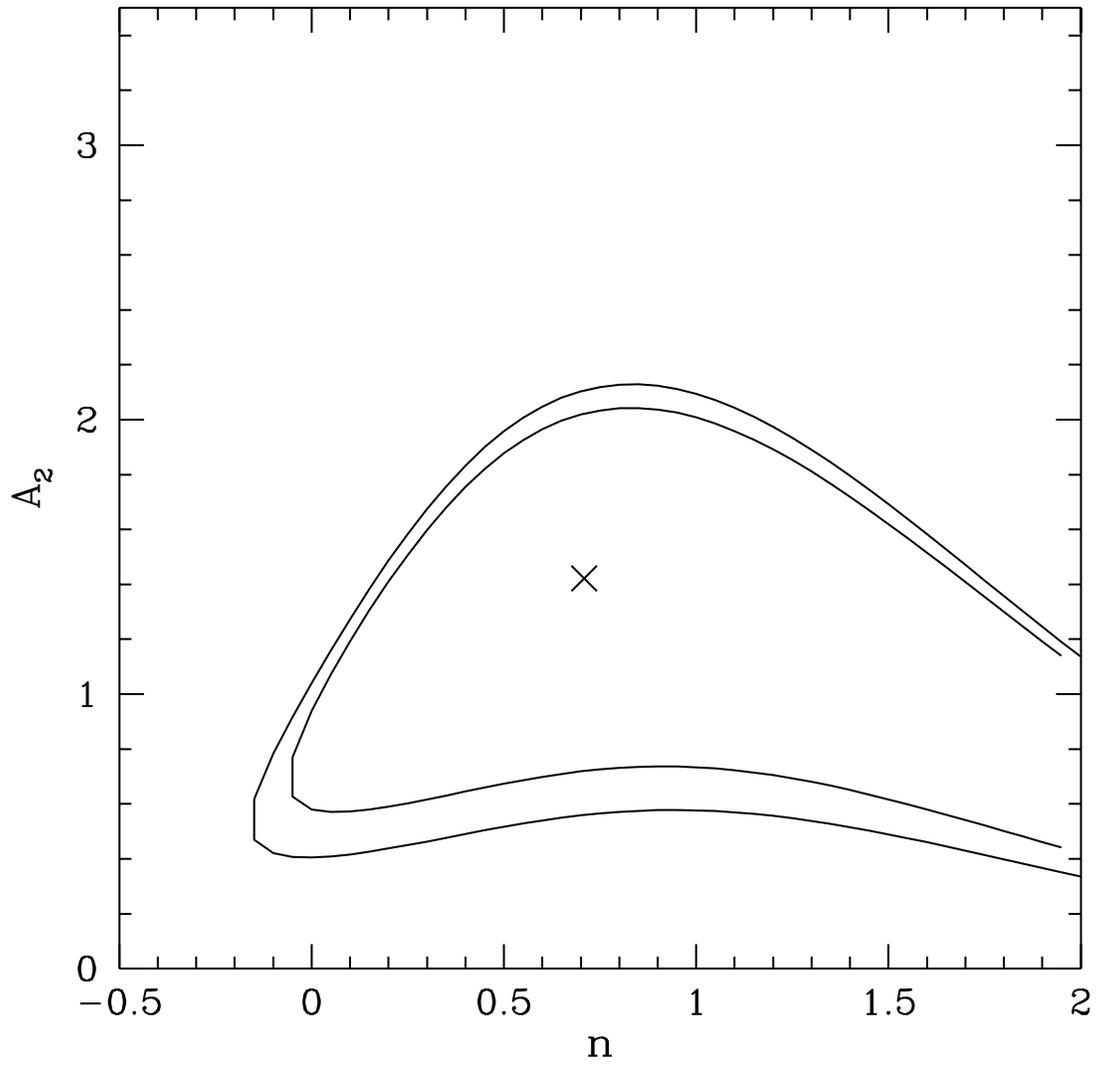

Figure 5:

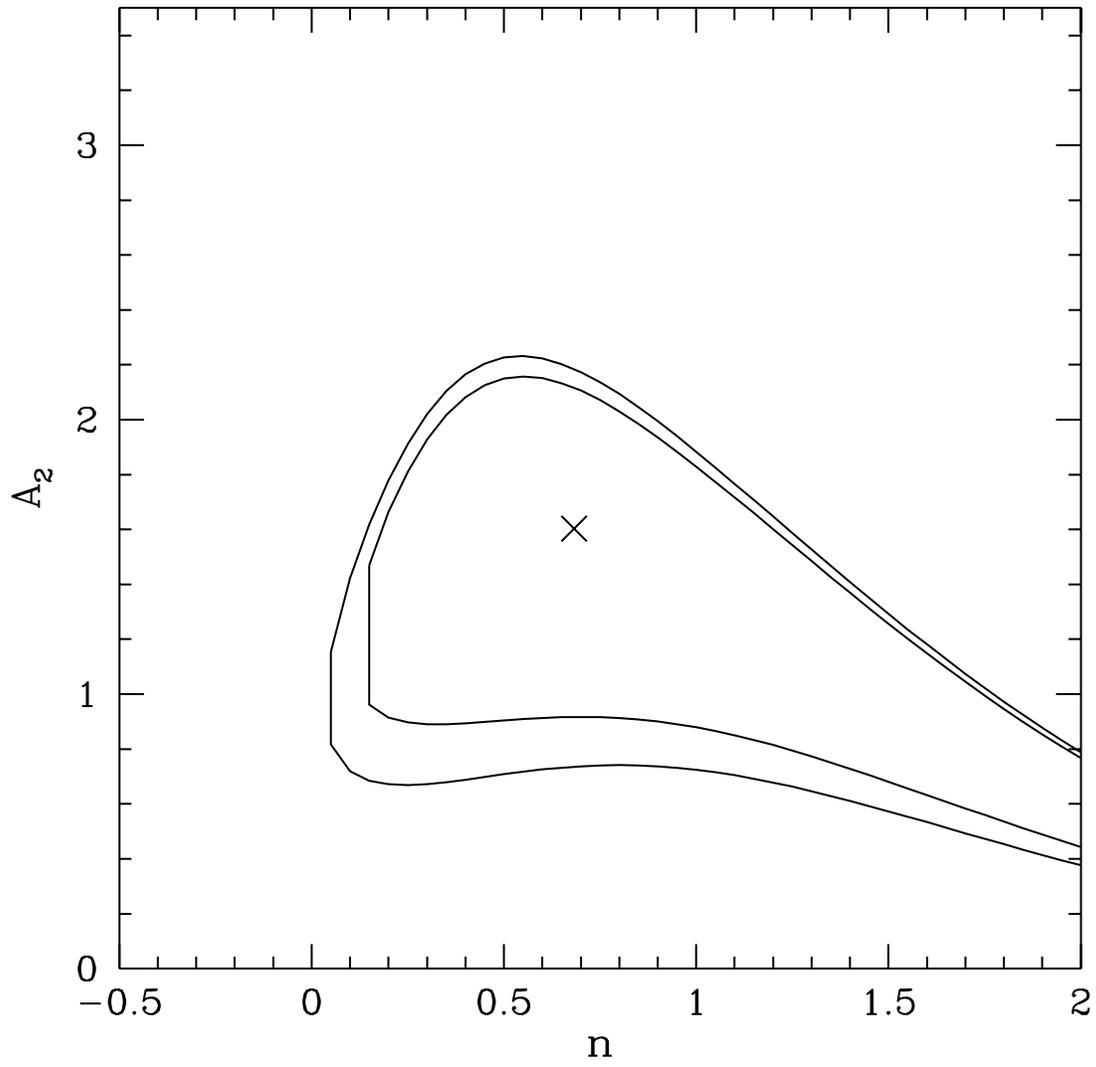

Figure 6:

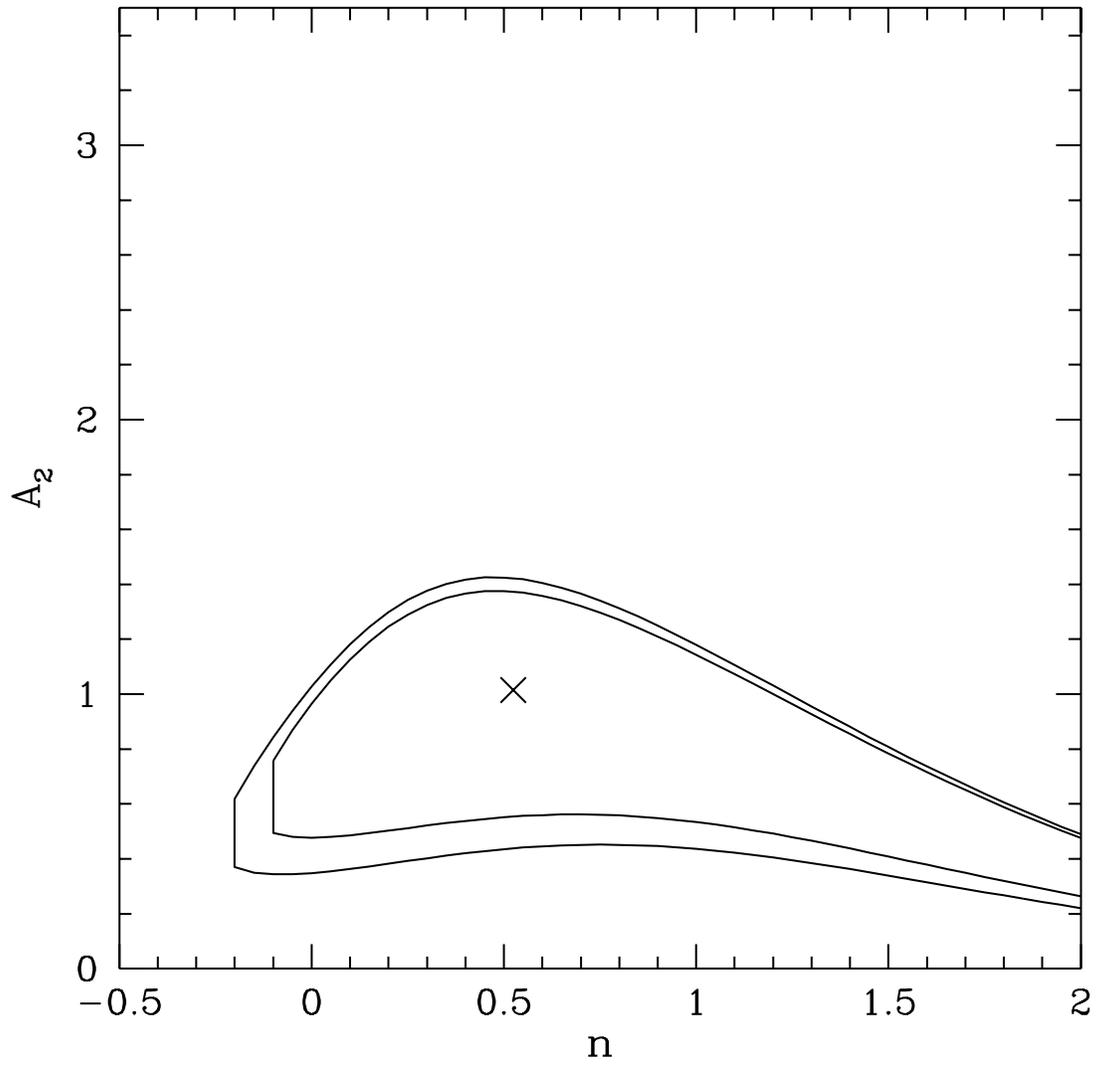

Figure 7:

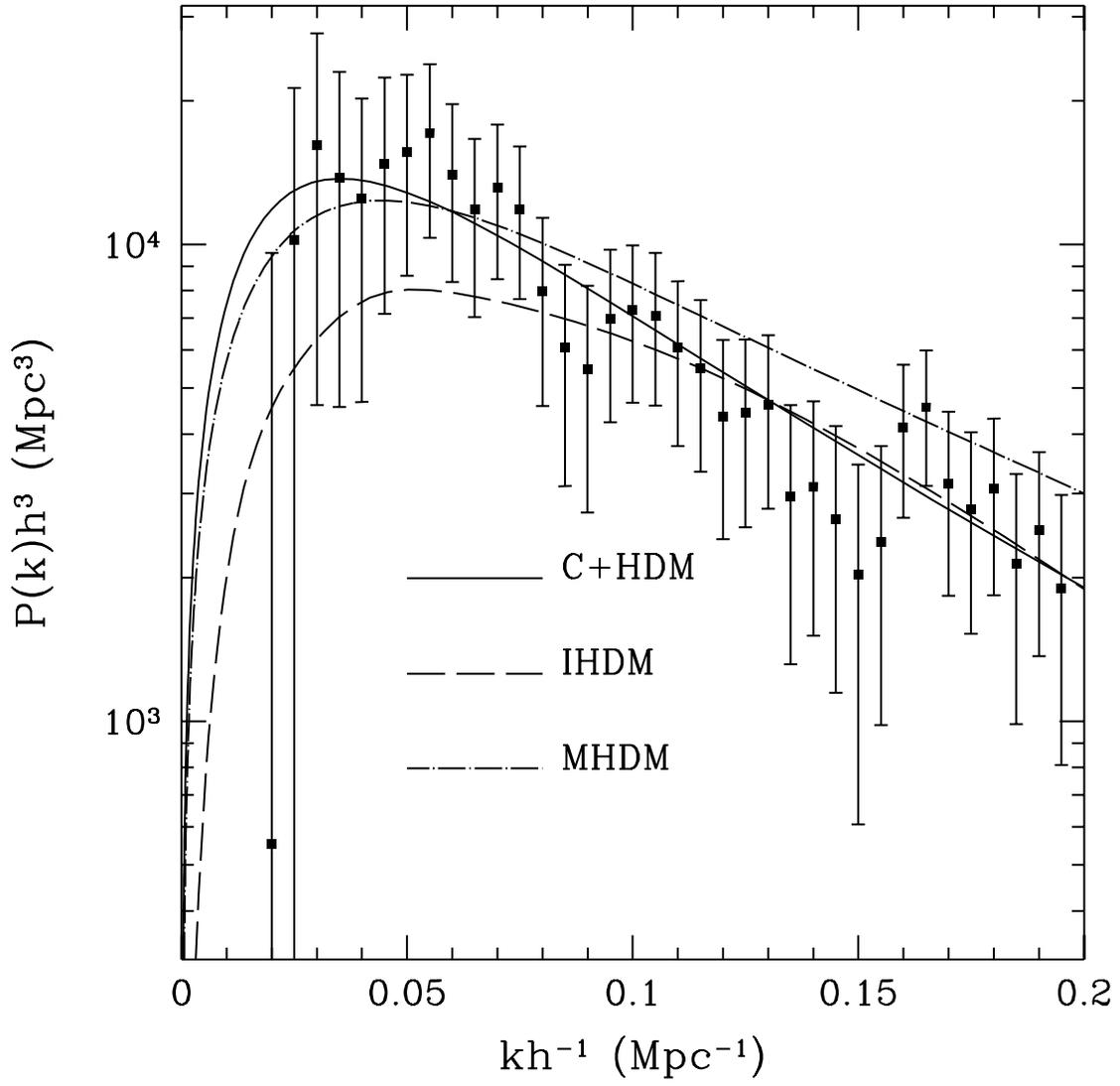

Figure 8:

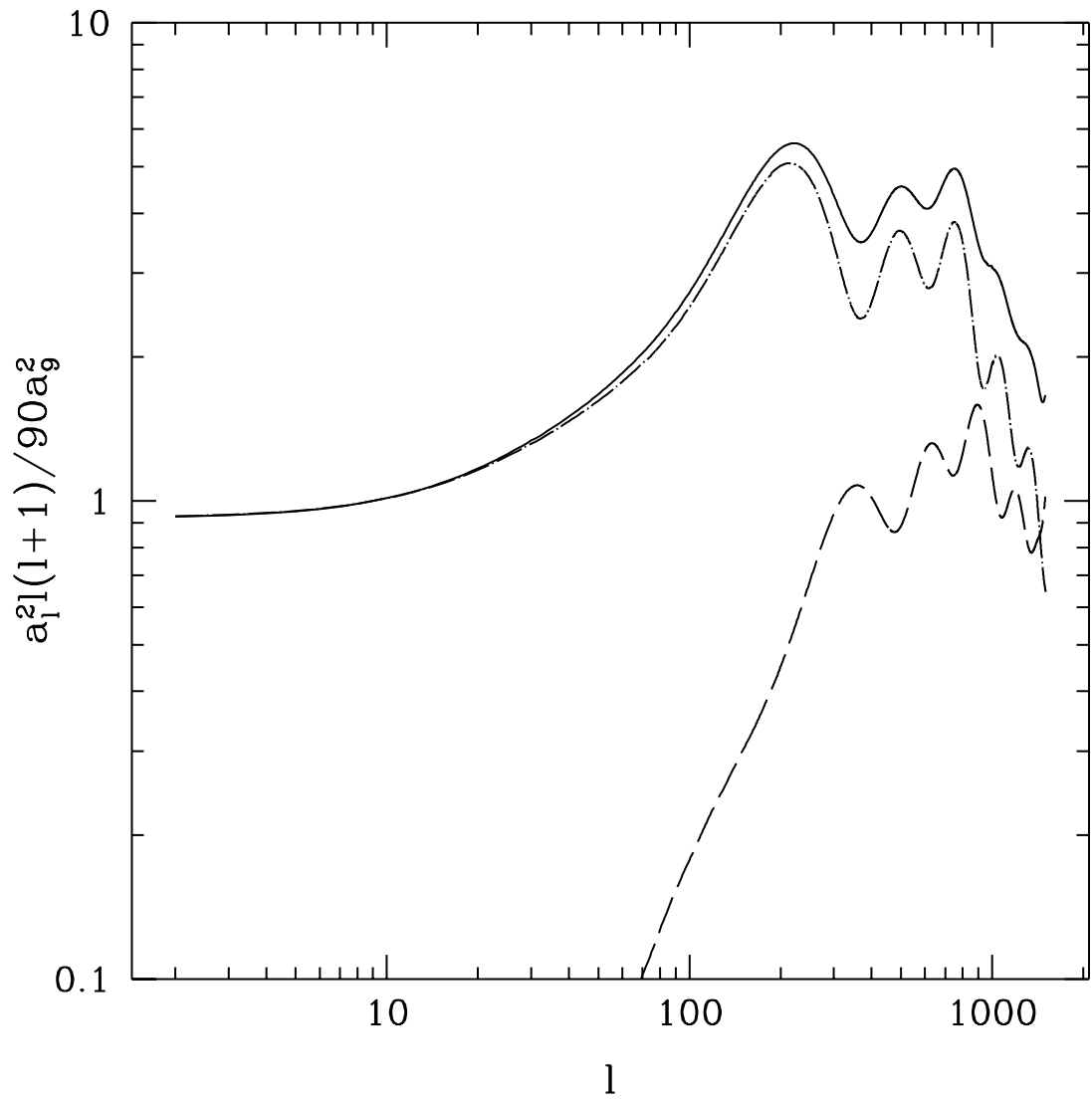

Figure 9:

Table 1:

| Coefficient | $z=0$ | $z=3.2$ | $z=4$ | $z=9$ |
|---:|---:|---:|---:|---:|
| $a_1$ | -1.658 | -2.284 | -1.601 | -2.051 |
| $a_2$ | 9.223(3) | 9.390(3) | 9.212(3) | 9.225(3) |
| $a_3$ | -1.774(5) | -1.718(5) | -1.656(5) | -1.4812(5) |
| $a_4$ | 3.100(6) | 2.943(6) | 2.852(6) | 2.420(6) |
| $a_5$ | -1.554(7) | -1.269(7) | -1.246(7) | -7.215(6) |
| $a_6$ | 3.720(7) | 1.739(7) | 1.945(7) | -9.235(6) |
| $a_7$ | 2.336(8) | 3.122(8) | 2.960(8) | 3.778(8) |
| $a_8$ | 5.388(-1) | 5.322(-1) | 5.338(-1) | 5.296(-1) |
| $a_9$ | 2.383(5) | 5.673(4) | 4.764(4) | 2.382(4) |
| $a_{10}$ | 1.283(-1) | 8.888(-2) | 8.577(-2) | 7.664(-2) |
| $a_{11}$ | -5.998(-1) | -7.272(-2) | 7.206(-4) | 3.045(-2) |
| $a_{12}$ | 3.888(-2) | 9.809(-3) | 4.437(-3) | 1.274(-3) |
| $a_{13}$ | 3.051 | 4.334 | 6.375 | 1.026(1) |
| $a_{14}$ | 2.003(-1) | 2.303(-1) | 2.839(-1) | 3.325(-1) |

Table 2:

| Coefficient | $z=0$ | $z=3.2$ | $z=4$ | $z=9$ |
|---|---:|---:|---:|---:|
| $a_1$ | 3.975(1) | 5.321(1) | 4.567(1) | 6.954(1) |
| $a_2$ | 8.330(2) | 4.552(2) | 8.306(2) | -3.230(2) |
| $a_3$ | -1.487(4) | -1.378(4) | -2.411(4) | 1.636(4) |
| $a_4$ | 3.791(5) | 7.804(5) | 8.224(5) | 7.328(5) |
| $a_5$ | -2.368(6) | -5.943(6) | -5.847(6) | -6.750(6) |
| $a_6$ | 7.555(6) | 1.982(7) | 1.861(7) | 2.547(7) |
| $a_7$ | -4.190(6) | -1.109(7) | -1.035(7) | -1.453(7) |
| $a_8$ | 1.087 | 1.053 | 1.073 | 9.697(-1) |
| $a_9$ | 1.002 | 1.001 | 1.001 | 1.001 |
| $a_{10}$ | 1.914 | 1.668 | 1.635 | 1.525 |
| $a_{11}$ | 8.899(-1) | 3.919(-1) | 3.696(-1) | 2.7212(-1) |
| $a_{12}$ | 2.339 | 7.210(-1) | 6.024(-1) | 4.178(-1) |
| $a_{13}$ | 1.321 | 1.365 | 1.419 | 1.254 |
| $a_{14}$ | 3.699(-1) | 3.224(-1) | 3.381(-1) | 2.800(-1) |
| $a_{15}$ | 1.605(1) | 4.268(1) | 3.93(1) | 5.984(1) |
| $a_{16}$ | 2.718(5) | 6.977(4) | 5.767(4) | 3.042(4) |

Table 3:

| Coefficient | |
|---|---:|
| $c_0$ | -13.66 |
| $c_1$ | -1.885 |
| $c_2$ | .2698 |
| $c_3$ | -4.238(-2) |
| $c_4$ | -1.537(-2) |
| $c_5$ | 4.856(-3) |
| $c_6$ | 1.356(-3) |